\documentclass[twoside,twocolumn,english,aps,prl]{revtex4}
\usepackage[T1]{fontenc}
\usepackage[latin1]{inputenc}
\usepackage{graphicx}
\usepackage{amssymb}

\makeatletter

\usepackage{graphicx}
\usepackage{amssymb}

\usepackage{graphicx}
\usepackage{amssymb}

\usepackage{graphicx}

\usepackage{graphicx}

\usepackage{babel}

\usepackage{babel}

\usepackage{babel}
\makeatother
\begin{document}

\title{Non-Fermi liquid and pairing in electron-doped cuprates}

\author{P. Krotkov$^{1}$ and Andrey V. Chubukov$^{1,2}$}

\affiliation{$^{1}$ Condensed Matter Theory Center, Department of Physics, University
of Maryland, College Park, Maryland, 20742 \\
 $^{2}$ Department of Physics, University of Wisconsin, 1150 University
Ave, Madison, WI 53706}

\begin{abstract}
We study the normal state and pairing instability in electron-doped
cuprates near optimal doping. We show that the fermionic self-energy
has a non-Fermi liquid form leading to peculiar frequency dependencies
of the conductivity and the Raman response. We solve the pairing problem
and demonstrate that $T_{c}$ is determined by the curvature of the
Fermi surface, and the pairing gap $\Delta(\mathbf{k},\omega)$ is
strongly non-monotonic along the Fermi surface. The normal state frequency
dependencies, the value of $T_{c}\sim10$K and the $\mathbf{k}-$dependence
of the gap agree with the experiment.

\end{abstract}
\maketitle
\newcommand{\sgn}{\mathop{\mathrm{sgn}}\nolimits}\newcommand{\const}{\mathop{\mathrm{const}}\nolimits} \newcommand{\ts}{\textstyle} \newcommand{\f}[1]{\mbox {\boldmath\(#1\)}}

Physics of high-temperature superconductors continues to attract considerable
attention. The past research primarily focused on hole-doped cuprates,
where a complex phase diagram was found with antiferromagnetic and
superconducting phases separated by a pseudogap region, and where
the normal (non-pseudogap) phase displays a prominent non-Fermi-liquid
behavior.

Recently there has been a surge of interest in \emph{electron-doped}
cuprates, Nd$_{2-x}$Ce$_{x}$CuO$_{4}$ and Pr$_{2-x}$Ce$_{x}$CuO$_{4}$.
Their phase diagram~\cite{el-pd} has a wider region of antiferromagnetism
(AFM); superconducting $T_{c}(x)$ forms a dome above the antiferromagnetic
quantum-critical point (QCP) \cite{dagan04}, much like in heavy fermion
materials~\cite{piers}; and reported pseudogap behavior~\cite{koitzsch03}
tracks the Neel temperature, and so is likely just a property of a
quasi-2D antiferromagnet \cite{zimmers05,Aiff03}. Optical data show~\cite{onose04}
the sharp charge-transfer gap near $2$eV at small $x$ melting away
near the optimal doping $x\sim0.15$, where the superconductivity
appears. We infer from these data and the experimental phase diagram
that near optimal doping the Mott physics is not crucial, and the
system behavior can be grasped by considering only low-energy electrons
that interact via collective bosons \cite{markiewicz03}. Electron-boson
models have been previously applied to hole-doped cuprates \cite{pines-scalapino,abanov03},
where Mott physics is more pronounced. They must therefore be tested
on electron-doped materials. In particular, they should explain a
much smaller $T_{c}$ in electron-doped cuprates despite almost the
same strength of the Hubbard interaction as in hole-dopes materials~\cite{millis04}.

\begin{figure}
\includegraphics[%
  width=0.7\columnwidth]{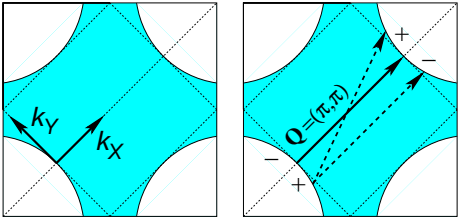}

\caption{Left: Fermi surface at the antiferromagnetic QCP with $2k_{F}=(\pi,\pi)$.
Diamond-shaped dashed lines bound the magnetic Brillouin zone. The
diagonal points of the FS (nodal points of the $d_{x^{2}-y^{2}}$-wave
gap) now become {}``hot''. Right: Graphic explanation of the attraction
in the $d_{x^{2}-y^{2}}$ channel: parts of the Fermi surface on the
same side of the zone diagonals are on average closer to the $\mathbf{Q}$
separation than the parts on the opposite sides, leading to attraction
in the $d_{x^{2}-y^{2}}$ channel (plus and minus are the signs of
the $d_{x^{2}-y^{2}}$ gap). \label{cap:sketch}}
\end{figure}

Confinement of the superconductivity to the vicinity of an antiferromagnetic
QCP implies that the most likely candidates for the pairing bosons
are the spin fluctuations with momenta near the antiferromagnetic
vector $\mathbf{Q=(\pi,\pi)}$. RPA studies suggest~\cite{onufrieva04}
that AFM appears close to the doping when $2k_{F}$ coincides with
$Q$, i.e. when the Fermi surface (FS) touches the boundary of the
AFM Brillouin zone at $\mathbf{k}_{F}=(\pi/2,\pi/2)$ (see Fig. \ref{cap:sketch}).
Near this point the free-fermion static susceptibility $\chi(\mathbf{Q})$
rapidly increases, leading to the Stoner instability.

Below we set $Q=2k_{F}$ and consider a spin-fluctuation model for
the $2k_{F}$ instability that involves fermions near $\mathbf{k}_{F}$
and their collective spin fluctuations near $\mathbf{Q}$. Such a
model qualitatively differs from the spin-fermion model for hole-doped
cuprates~\cite{abanov03}, where relevant fermionic states (hot spots)
are located far from the zone diagonals. In that case, the spin-fluctuation
exchange led to the $d_{x^{2}-y^{2}}$-wave pairing~\cite{pines-scalapino}
with the gap maxima at the hot spots~\cite{abanov03}. When the dominant
interaction is between the nodal fermions, it is \emph{a priori} unclear
to what degree the $d_{x^{2}-y^{2}}$-wave pairing survives.

\begin{figure}
\includegraphics[%
  width=0.7\columnwidth,
  keepaspectratio]{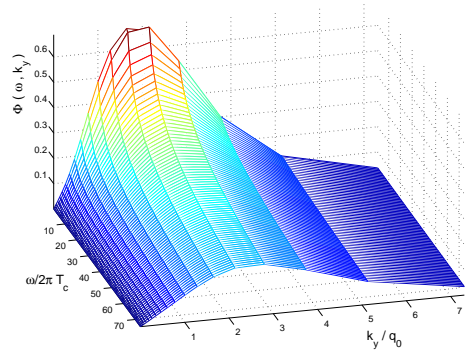}

\caption{The pairing vertex $\Phi(k_{y},\omega_{n})$ for $r=0.05$ vs $\omega_{n}/(2\pi T_{c})$
and $k_{y}/q_{0}$ ($q_{0}$ is defined in the text). Observe that
$\Phi(k_{y},\omega_{n})$ is non-monotonic in $k_{y}$.\label{cap:gap}}
\end{figure}

In this paper we address this and other issues for $2k_{F}$ instability.
We confirm earlier result~\cite{altshuler95} that the interaction
with low-energy spin fluctuations destroys Fermi liquid behavior for
diagonal fermions. We show that this causes a peculiar behavior of
the optical conductivity $\sigma(\omega)\sim\omega^{-0.64}$ over
a wide frequency range, and a nearly frequency-independent Raman response
in the $B_{2g}$ channel. Both of the results agree with the experiment~\cite{homes,quazilbash05}.
We then use the normal state results as an input and consider for
the first time the spin-mediated pairing near $2k_{F}$ instability.
We demonstrate that $T_{c}$ is finite at the QCP and the pairing
is dominated by non Fermi-liquid fermions. We show that the pairing
problem is very peculiar as now the curvature of the fermionic dispersion
plays a major role. We found $T_{c}\sim10$K for the same parameters
as for hole-doped materials. We also found non-monotonic variation
of the $d-$wave gap $\Delta(k,\omega)$ along the FS (see Fig.\ref{cap:gap}).
Non-monotonic variation of the gap agrees with the numerical studies~\cite{num}
and with the earlier BCS-type study~\cite{yakovenko04}, but our
solution is very different from the BCS one. The non-monotonic $d-$wave
gap has been detected experimentally at smaller $x$ and has been
traced to the location of the hot spots~\cite{blumberg02,sato}.
Our results show that near a magnetic QCP, the positions of the gap
maxima deviate from the hot spots, and remain fixed even when the
hot spots merge and then disappear.

We now provide the details of the calculations. A spin fermion model
describes low-energy fermions $c_{\mathbf{k}}$ interacting with their
collective bosonic fluctuations in the spin channel $\mathbf{S}_{\mathbf{q}}$:
$\mathcal{H}_{\mathrm{int}}=g\sum_{\mathbf{k},\mathbf{q}}c_{\mathbf{k}+\mathbf{q},\alpha}\f\sigma_{\alpha\beta}c_{\mathbf{k}\beta}\mathbf{S}_{\mathbf{q}}$
(for internal consistency of the model it is required that $g\lesssim E_{F}$~\cite{abanov03}).
The spin subsystem is described by the static spin susceptibility
which, as before, we assume to have the Ornstein-Zernike form $\chi_{0}(\mathbf{q})=\chi_{0}/[\xi^{-2}+(\mathbf{q}-\mathbf{Q})^{2}]$.
Interacting fermions are located near $\mathbf{k}_{F}$ and $\mathbf{k}_{F}+\mathbf{Q}$,
and have almost antiparallel velocities: \begin{equation}
\epsilon_{\mathbf{k}}\approx v_{F}\Delta k_{x}+\beta^{2}\Delta k_{y}^{2},\quad\epsilon_{\mathbf{k}+\mathbf{Q}}\approx-v_{F}\Delta k_{x}+\beta^{2}\Delta k_{y}^{2},\label{eq:ep}\end{equation}
 where $\Delta\mathbf{k}=\mathbf{k}-\mathbf{k}_{F}$. Factor $\beta$
accounts for the curvature of the Fermi line. The strategy is to solve
self-consistently for the normal state bosonic $\Pi(\mathbf{q},\Omega)$
and fermionic $\Sigma(\mathbf{k},\omega)$ self-energies, and then
use the full normal state propagators as input for the pairing problem.
We define the fermionic propagator as $G(\mathbf{k},\omega)^{-1}=i(\omega+\Sigma(\mathbf{k},\omega))-\epsilon_{\mathbf{k}}$
and bosonic as $\chi(\mathbf{q},\omega)^{-1}=\chi_{0}^{-1}(\mathbf{q})+2[\Pi(\mathbf{q},\omega)+\Pi(2\mathbf{Q}-\mathbf{q},\omega)]$.

\paragraph{Normal state}

Fermionic and bosonic self-energies in the normal state have been
obtained by Altshuler et al.~\cite{altshuler95} within Eliashberg
theory, and we just quote their result: at the $2k_{F}$ antiferromagnetic
QCP and $k=k_{F}$ along the diagonal, \begin{equation}
\Pi(\mathbf{Q},\Omega)=\frac{\bar{g}}{2\pi v_{F}\beta}\sqrt{|\Omega|},\;\;\;\Sigma(\mathbf{k}_{F},\omega)=\omega_{0}^{1/4}|\omega|^{3/4},\label{eq:PiBare}\end{equation}
 where $\bar{g}=g^{2}\chi_{0}$, and $\omega_{0}=(\bar{g}\beta/\pi Nv_{F})^{2}$.
Both self-energies differ from the case when the hot spots are far
from zone diagonals. E.g., $\Pi(\mathbf{q},\Omega)$ scales as $|\Omega|^{1/2}$
instead of the conventional Landau damping form $\Pi(\mathbf{q},\Omega)\propto|\Omega|$.
This is a consequence of antiparallel velocities at $\mathbf{k}_{F}$
and $\mathbf{k}_{F}+\mathbf{Q}$ (observe that the damping rate $\Pi(q,\Omega)\propto1/\beta$
does not diverge only because the curvature $\beta\ne0$). Eq. \ref{eq:PiBare}
is strictly speaking valid for $\omega>\omega_{0}$. For $\omega<\omega_{0}$
expressions for the self-energies are more involved~\cite{altshuler95},
and one has to sum up series of logarithms to obtain the proper exponents.
Altshuler et al. found $\Sigma(\omega)\sim\omega_{0}^{\alpha}\omega^{1-\alpha}$,
where $\alpha\sim0.14$. We confirmed their result. This low-frequency
behavior, however, is less important for practical purposes, since
we found that $\omega_{0}\sim10$meV is quite small~\cite{comm2}.
Relevant physics, including the pairing, then mostly comes from frequencies
$\omega>\omega_{0}$, where Eqs. (\ref{eq:PiBare}) are valid.

For the FS momenta away from the diagonal ($k_{y}\neq0$, see Fig.
\ref{cap:sketch}), the non-Fermi liquid form of the self-energy (\ref{eq:PiBare})
survives up to $k_{y,\mathrm{max}}\sim\omega^{1/4}$, and crosses
over to the Fermi liquid form at larger $k_{y}$. At the same time,
for $\mathbf{k}$ transverse to the FS the self-energy is regular
already at $k_{y}=0$: $\Sigma(k_{x},0)\propto k_{x}$.

\begin{figure}
\includegraphics[%
  width=1\columnwidth]{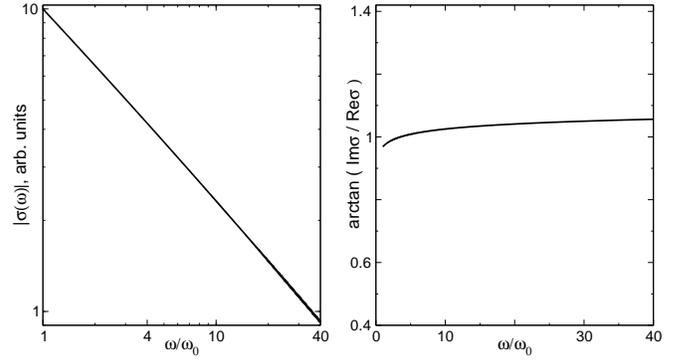}

\caption{Normal state conductivity as a function of frequency shows a scaling
behavior for $\omega_{0}<\omega<40\omega_{0}$. Left: $|\sigma(\omega)|$,
the behavior is indistinguishable from the $\sigma(\omega)\sim\omega^{-0.64}$
dependence. Right: $\arctan\Im\sigma(\omega)/\Re\sigma(\omega)$,
an almost constant value means that both $\Im\sigma(\omega)$ and
$\Re\sigma(\omega)$ scale as $\omega^{-0.64}$. \label{cap:conductivity}}
\end{figure}

\paragraph{Optical conductivity}

We used the result for the self-energy to compute complex conductivity
$\sigma(\omega)$ from the Kubo formula appropriate when the self-energy
mainly depends on frequency. We verified that the dominant contribution
to $\sigma(\omega)$ comes from nodal regions where $\Sigma(\omega)$
has a non-Fermi-liquid form \cite{comm-1}. At very small frequencies,
$\omega<\omega_{0}$, $\Sigma(\omega)>\omega$, and both $\Im\sigma$
and $\Re\sigma$ have the same power-law behavior: $\sigma(\omega)\propto\omega^{-1+2\alpha}\sim\omega^{-0.7}$.
At larger frequencies, $\Sigma(\omega)<\omega$, and one should not
generally expect $\Re\sigma$ and $\Im\sigma$ to scale with each
other. Surprisingly, the scaling behavior extends, with almost the
same power, $\sigma(\omega)\propto\omega^{-0.64}$ up to $\sim40\omega_{0}$
(see Fig. \ref{cap:conductivity}). Such power-law behavior is not
a sign of a true scaling, but rather a consequence of the fact that
$\Sigma(\omega)/\omega\propto(\omega_{0}/\omega)^{1/4}$ is a slow
decaying function. The $\omega^{-\gamma}$ behavior of conductivity
with $\gamma\approx0.68$ has been observed in Pr$_{1.85}$Ce$_{0.15}$CuO$_{4}$
below $400$meV ~\cite{homes}. Both the exponent and the experimental
frequency range are quite consistent with our results. Note that a
very similar behavior of the conductivity at intermediate energies
has been earlier observed in hole-doped cuprates~\cite{marel}.

\paragraph{Raman response}

We also considered the Raman scattering in the $B_{2g}$ channel.
Raman intensity is proportional to the imaginary part of the polarization
bubble weighted with the geometry dependent Raman form factors $\gamma_{k}$.
In the $B_{2g}$ geometry, $\gamma(\mathbf{k)}\propto\sin k_{x}\sin k_{y}$,
such that the signal comes primarily from around the diagonal direction.
Unlike the current vertex for conductivity, the Raman vertex is renormalized
by the interaction even when $\Sigma(\omega)$ depends only on frequency.
Because $\gamma(\mathbf{k})$ is a scalar and doesn't change sign
under $\mathbf{k}\rightarrow\mathbf{k}+\mathbf{Q}$, we can approximate
it by a constant $\gamma(\mathbf{k}_{F})$. The renormalization of
the Raman vertex then coincides with that of the density vertex, and
is related to the self-energy by the Ward identity: the full $\gamma_{\mathrm{full}}(\mathbf{k},\omega)=\gamma(\mathbf{k})(1+\partial_{\omega}\Sigma(\omega))$.
Evaluating the Raman bubble and substituting $\gamma_{\mathrm{full}}$
into it, we then obtain \begin{equation}
R_{B_{2g}}(\omega)\sim k_{y,\mathrm{max}}(\omega)\left(1+\partial_{\omega}\Sigma(\omega)\right)^{2}\frac{\omega}{\omega+\Sigma(\omega)}\end{equation}
 Substituting the forms of the self-energy and $k_{y,\mathrm{max}}(\omega)\sim\omega^{\alpha}$
for $\omega<\omega_{0}$, $k_{y,\mathrm{max}}(\omega)\sim\omega^{1/4}$
for $\omega>\omega_{0}$, we find that at $T=0$, $R_{B_{2g}}(\omega)$
is flat: it is a constant at $\omega<\omega_{0}$, and very slowly
crosses over to the $(\omega/\omega_{0})^{-1/4}$ behavior at $\omega>\omega_{0}$.
The flat form of the Raman intensity at frequencies $\omega>\omega_{0}$
is consistent with the experimental data~\cite{quazilbash05}. The
experimental behavior at $\omega\leq\omega_{0}$ is dominated by temperature
effects which we do not consider.

\paragraph{Pairing instability}

We now apply normal state results to study the pairing instability
near a magnetic QCP. As we said in the introduction, it is \emph{a
priori} unclear whether $d-$wave pairing survives when the hot spots
merge, as the strongest pairing interaction involves quasiparticles
for which the $d_{x^{2}-y^{2}}$ superconducting gap vanishes. However,
separation between near-nodal fermions with opposite signs of the
$d_{x^{2}-y^{2}}$ gap is on average closer to $\mathbf{Q}$ than
that between fermions with the same sign of the gap (see Fig. \ref{cap:sketch}
b). So the spin-fluctuation exchange still leads to attraction in
the $d-$wave channel. Furthermore, as the gap equation relates $\Delta(\mathbf{k})$
near $\mathbf{k}_{F}$ and $\mathbf{k}_{F}+\mathbf{Q}$, where both
gaps are linear in the deviation from the nodal direction, the gap
equation becomes an equation on the slope of $\Delta(\mathbf{k})$
and does not contain any extra smallness.

\begin{figure}
\includegraphics[%
  width=0.7\columnwidth,
  keepaspectratio]{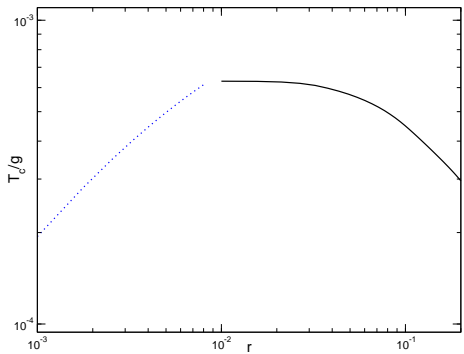}

\caption{Pairing instability temperature $T_{c}$ in units of the coupling
constant $\overline{g}$ vs. the curvature parameter $r=\bar{g}\beta^{2}/\pi v_{F}^{2}$.
Solid line gives the result of the numeric solution, which requires
increasingly long computational time as $r\rightarrow0$. Dashed line
is the analytic small-$r$ form $T_{c}=0.95\bar{g}r/(1+5r^{2/5})^{4}$.
\label{cap:Tc}}
\end{figure}

The linearized gap equation at QCP is obtained using Eliashberg technique
for collective-mode mediated pairing~\cite{abanov03}. We verified
that the dominant contribution to the pairing comes from $\omega>\omega_{0}$,
where $\Sigma(\omega)$ can be approximated by the $\omega^{3/4}$
form. To single out the role of the curvature, we rewrite $\omega_{0}$
as $\omega_{0}=\omega_{1}r$, where $\omega_{1}=\bar{g}/(4\pi)$ accounts
solely for the interaction, and the dimensionless $r=\bar{g}\beta^{2}/\pi v_{F}^{2}\sim\bar{g}/E_{F}$
parameterizes the effect of the curvature. Introducing dimensionless
variables $k/q_{0}$, $\omega/\omega_{1}$, where $q_{0}=\bar{g}/(\pi v_{F}r^{1/4})\sim k_{F}(\bar{g}/E_{F})^{3/4}$,
and assuming that the gap has $d_{x^{2}-y^{2}}$ symmetry, i.e., $\Delta(\mathbf{k}_{F})=-\Delta(\mathbf{k}_{F}+\mathbf{Q})$,
we obtain an integral equation for the pairing vertex $\Phi(k_{y},\omega_{n})$
near a single hot spot: \begin{eqnarray}
\Phi(k_{y},\omega_{n}) & = & \frac{3}{4}T\sum_{\omega_{m}}\frac{r^{1/4}}{|\omega_{m}|+r^{1/4}|\omega_{m}|^{3/4}}\label{eq:inteq}\\
 & \times & \int\frac{\Phi(k_{y}',\omega_{m})dk_{y}'}{(k_{y}-k_{y}')^{2}+r^{3/2}(k_{y}^{2}+k'_{y}{}^{2})^{2}+\sqrt{|\omega_{n-m}|}}.\nonumber \end{eqnarray}
 The momentum dependence in the kernel in (\ref{eq:inteq}) comes
from $(\mathbf{k}-\mathbf{k}')^{2}$ after the substitution $k_{x}=\beta^{2}k_{y}^{2}/v_{F}$
and $k'_{x}=-\beta^{2}k'_{y}{}^{2}/v_{F}$. The pairing vertex is
related to $\Delta$ as $\Phi=\Delta\omega/(\omega+\Sigma(\omega))$.
A $d_{x^{2}-y^{2}}$-wave solution of (\ref{eq:inteq}) is antisymmetric
in $k_{y}$ and symmetric in $\omega$~\cite{comm4}.

At vanishing $r$, one can neglect the curvature of the fermionic
dispersion in the gap equation (the $r^{3/2}$ term in the denominator
in (\ref{eq:inteq})). Without this term, the kernel in (\ref{eq:inteq})
is non-Fredholmian, and its antisymmetric-in-$k$ solution is a monotonic
$\Phi(k_{y},\omega_{n})=k_{y}\Phi(\omega_{n})$. Substituting this
form into (\ref{eq:inteq}), we obtain for $\Phi(\omega_{n})$: \begin{equation}
\Phi(\omega_{n})=\frac{3}{4}\pi T\sum_{m\ne n}\frac{\omega_{0}^{1/4}\Phi(\omega_{m})}{|\omega_{n-m}|^{1/4}\left(|\omega_{m}|+\omega_{0}^{1/4}|\omega_{m}|^{3/4}\right)}.\label{eq:integ_gamma}\end{equation}
 Expectedly (since we neglected the curvature), this equation is the
same as in the Eliashberg theory for the pairing, mediated by a bosonic
susceptibility $\chi(\omega)\propto\omega^{-1/4}$ \cite{aacy}. Numerical
solution of (\ref{eq:integ_gamma}) gives $T_{c1}\approx6\omega_{0}=6\omega_{1}r$,
i.e. $T_{c}\sim r$. We emphasize that the $r\rightarrow0$ limit
is not weak coupling (self-energy cannot be neglected) despite the
factor $r^{1/4}$ in the numerator of (\ref{eq:inteq}).

When the curvature of the fermionic dispersion is included, Eq. (\ref{eq:inteq})
becomes Fredholmian, and a numerical solution can be obtained using
the Nystrom discretization method \cite{nr}. We found that the effect
of the curvature on $T_{c}$ is very strong: above $r>0.001$, the
actual $T_{c}$ rapidly becomes much smaller than $T_{c1}$. For $r\sim0.1$,
$T_{c}\sim0.01T_{c1}$. We plotted $T_{c}(r)$ in Fig. \ref{cap:Tc}.
We see that over a wide range $0.01<r<0.1$, $T_{c}\sim0.006\omega_{1}\approx0.0005\bar{g}$
forms a plateau and weakly depends on $r$. Using the same $\bar{g}\sim1.6$eV
as for hole-doped cuprates \cite{abanov02}, we obtained in this range
$T_{c}\sim10$K. Using further the previous estimate of $\omega_{0}=\omega_{1}r\sim10$meV
we obtain $r\sim0.08$, which is within the region where $T_{c}$
is almost a constant. Out $T_{c}\sim10K$ at a magnetic QCP is in
agreement with the experiment \cite{el-pd,quazilbash05,zimmers05}.
Note that the experimental $T_{c}$ increases in the antiferromagnetic
phase up to $20$K, and only then drops at smaller dopings.

The huge discrepancy between $T_{c}$ and $T_{c1}$ can also be understood
analytically, by expanding $T_{c}$ in $r$ beyond the $O(r)$ term.
We found that the expansion is rather non-trivial and holds in fractional
powers of $r^{2/5}$: $T_{c}\approx T_{c1}/(1+ar^{2/5})^{4}$. The
prefactor $a>0$ cannot be obtained analytically, but the fit to the
numerical data gives $a\sim5$. As $(1+5r^{2/5})^{4}$ rapidly increases
with $r$ (it is $\sim80$ for $r=0.1$), this explains why $T_{c}$
is much smaller than $T_{c1}$.

We also found that at finite $r$ the gap is a non-monotonic function
of $k_{y}$: it is linear at small $k_{y}$, passes through a maximum
at a $k_{0}$ and then falls off (see Fig. \ref{cap:gap}). Position
of the maximum depends on $r$ and appears uncorrelated with the positions
of hot spots, which at the QCP are on the zone diagonals. At small
$r$ we found analytically that $k_{\mathrm{0}}(r)\sim k_{F}r^{3/5}\sim k_{F}(\bar{g}/E_{F})^{3/5}$.
As $r$ is small, $k_{\mathrm{0}}<k_{F}$ and the gap is confined
to the vicinity of the zone diagonals.

At deviations from the QCP towards larger dopings, into a paramagnetic
phase, $T_{c}$ decreases and eventually disappears. The value of
$k_{\mathrm{0}}$, however, does not track the decrease of $T_{c}$,
i.e., the $d-$wave gap extends over a finite momentum range along
the FS even in the overdoped materials. At deviations into the antiferromagnetic
phase, the FS evolves into hole and electron pockets, and the locations
of $k_{\mathrm{0}}$ gradually approach the hot spots.

Our results for $T_{c}$ and the gap survive even when the magnetic
correlation length $\xi$ remains finite at the doping where $2k_{F}=Q$,
i.e., when antiferromagnetic QCP is shifted to lower dopings, when
the hot spots are away from zone diagonals. We found that the modifications
of our results are small as long as $T_{c}>J\xi^{-2}$ where $J$
is the exchange interaction.

\paragraph{Summary }

We considered the normal state properties and pairing near a $2k_{F}$
antiferromagnetic QCP and applied the results to electron-doped cuprates
at optimal doping. We found that the breakdown of the Fermi-liquid
description at QCP leads to peculiar frequency dependences of the
conductivity and the $B_{2g}$ Raman response. We found that the superconducting
$T_{c}$ remains finite at the QCP and is of order $10$K for the
same spin-fermion coupling as for hole-doped cuprates. The pairing
gap at QCP has $d_{x^{2}-y^{2}}$ symmetry, but is highly anisotropic
and confined to momenta near the zone diagonals.

We acknowledge useful discussions with G. Blumberg, R. Greene, D.
Drew, C. Homes, A. Millis, M. Norman, M. M. Qazilbash, and V. Yakovenko.
We thank G. Blumberg, A. Millis and M. M. Qazilbash for the comments
on the manuscript. The research is supported by Condensed Matter Theory
Center at UMD (P.K, A.C) and by NSF DMR 0240238 (A.C.).


\begin{thebibliography}{15}
\bibitem{el-pd}Y Tokura, H. Takagi, S. Uchida , Nature \textbf{337}, 345 (1989). 
\bibitem{dagan04}Y. Dagan et al., Phys. Rev. Lett. \textbf{92}, 167001 (2004).
\bibitem{piers}P. Coleman and C. Pepin, Acta Physica Polonica B \textbf{34}, 691
(2003). 
\bibitem{koitzsch03}A. Koitzsch et al., \prb \textbf{67}, 184522 (2003). 
\bibitem{zimmers05}A. Zimmers et al., Europhys. Lett. \textbf{70}, 225 (2005). 
\bibitem{Aiff03}L. Alff et al., Nature \textbf{422}, 698 (2003). 
\bibitem{onose04}Y. Onose et al., Phys. Rev. B \textbf{69}, 024504 (2004). 
\bibitem{markiewicz03}R.S. Markiewicz, in \emph{Intrinsic Multiscale Structure and Dynamics
in Complex Electronic Oxides}, ed. by A.R. Bishop, et al., World Scientific
(2003). 
\bibitem{pines-scalapino}D.J. Scalapino, Phys. Rep. \textbf{250}, 329 (1995); P. Monthoux and
D. Pines, Phys. Rev. B \textbf{47}, 6069 (1993). 
\bibitem{abanov03}A. Abanov et al., Adv. Phys. \textbf{52}, 119 (2003). 
\bibitem{millis04}A.J. Millis et al., cond-mat/0411172 (unpublished). 
\bibitem{onufrieva04}F. Onufrieva and P. Pfeuty, Phys. Rev. Lett. \textbf{92}, 247003 (2004). 
\bibitem{altshuler95}B.L. Altshuler, L.B. Ioffe, A.J. Millis, Phys. Rev. B \textbf{52},
5563 (1995). The self-energy at finite temperature has been obtained
by O. Tchernyshyov and A.V. Chubukov (unpublished). The case of a
nested Fermi surface has been analyzed by A. Virosztek and J. Ruvalds,
Phys. Rev. B \textbf{42}, 4064 (1990). 
\bibitem{quazilbash05}M.M. Qazilbash et al., cond-mat/0510098, to be published in PRB. 
\bibitem{homes}C. Homes, unpublished 
\bibitem{num}K. Yoshimura and Hirashima, J. Phys. Soc. Jpn. \textbf{74}, 712 (2005)
and references therein. 
\bibitem{yakovenko04}V.A. Khodel et al., Phys. Rev. B \textbf{69}, 144501 (2004). 
\bibitem{blumberg02}G. Blumberg et al., Phys. Rev. Lett. \textbf{88}, 107002 (2002); ibid.
\textbf{90}, 149702 (2003). 
\bibitem{sato}H. Matsui et al., Phys. Rev. Lett. \textbf{95}, 017003 (2005). 
\bibitem{comm2}We used the same parameters as for hole-doped materials ~\protect\cite{abanov02}
$g=0.7$eV, $\chi_{0}\sim3$eV ($\bar{g}=g^{2}\chi_{0}\sim1.6eV$),
and $t-t^{\prime}$ model that fits the ARPES data for Bi$2212$ to
get $v_{F}$ and $\beta$. Note that an analogous scale in hole-doped
materials is $\sim250$meV~\protect\cite{abanov03}. 
\bibitem{comm-1}This is valid at $T=0$ and finite frequencies, when short-circuiting
of the conductivity (R. Hlubina, T.M. Rice, \prb \textbf{51}, 9253
(1995)) is not important. 
\bibitem{marel}D. van der Marel et al., Nature \textbf{425}, 271 (2003). 
\bibitem{comm4}A symmetric in $k_{y}$ solution of (\ref{eq:inteq}) that would correspond
to a $p-$wave gap stipulates a different overall sign before the
kernel; and an odd-frequency $d-$wave solution has lower $T_{c}$
than the even-frequency one. 
\bibitem{aacy}Ar. Abanov et al. (unpublished). 
\bibitem{nr}W.H. Press et al, \emph{Numerical Recipes: The Art of Scientific Computing},
New York, 2002. 
\bibitem{abanov02}Ar. Abanov et al., Phys. Rev. Lett, \textbf{89}, 177002 (2002). 
\bibitem{chub-norman}A.V. Chubukov, M.R. Norman, \prb  \textbf{70}, 174505 (2004). 
\end{thebibliography}
\end{document}